# Statistical Studies of Long – Period Variable Stars in Odessa

Larisa S. Kudashkina[1], Ivan L. Andronov[1],
Vladislava I. Marsakova[2], Lidia L. Chinarova[3]

[1]Department "High and Applied Mathematics", Odessa National Maritime University, Odessa, Ukraine
[2]Department of Astronomy, Odessa National University, Odessa, Ukraine
[3]Astronomical Observatory, Odessa National University, Odessa, Ukraine

Abstract

The studies of pulsating variable stars are traditional subjects of astronomers in Odessa. In the last half of the 20th century, the studies of the physical variable stars were the topics of 15 PhD theses of the collaborators of the Odessa State (now National) University.

Continuing the tradition of studies of long-period variable stars in the Astronomical Observatory of Odessa University, we try to conduct a detailed classification of the Mira-type stars, semi-regular variable stars and also find out the location of symbiotic stars among other long-period variables in their evolution to the planetary nebulae, using the compiled material from own observations and that of other authors.

For the research, we have used observations from the databases of the French Association of Variable Stars Observers (AFOEV) and Variable Stars Observers League of Japan (VSOLJ), which allow study of the variability of these stars during the time interval of about 100 years. Some stars were studied using the observations of the American Association of Variable Star Observers (AAVSO).

Modern research of long-period variable stars in Odessa is conducted in several directions:
1. The analysis of correlations between the photometric parameters.
2. The study of changes on time of individual characteristics of the light curves.
3. The study of variability of the objects which are transitional between the Mira-type and the semi-regular variables.

The conclusions are discussed.

The studies of pulsating variable stars are traditional subjects of astronomers in Odessa. In the last half of the 20th century, the studies of the physical variable stars were the topics of 15 PhD theses of the Odessa State (now National) University (Karetnikov, 1996).

Let's list the basic, the most fundamental works.

The first was the PhD thesis by E.P. Strelkova "The study of the change of the color indexes of the Mira-type long-period variable stars" (1956). The main results of this work are described in the article "The color indexes 13 Mira-type long-period variable stars" (Strelkova, 1956). Already from this work it is clear how it is difficult to classify variable stars like Mira using a small number of characteristics (for example, the period, magnitudes at extrema and color indexes). Four types of color index changes corresponding to the curve were selected. However, almost all of these types can be determined during several pulsating cycles of an individual star.

The second fundamental work that we note here, was defended in 1977, A.G. Nudzenko: "Changeability of periods of long-period variable stars". The main results were published in the article "Study of the (O-C)-diagrams of the Mira-type stars" (Nudzhenko, 1974). Just 238 Mira-type stars were studied. The considered dependences are $|\Delta P/P|$ versus P, $\Delta E$ versus P and $|\Delta P/P|$ versus $\Delta E$, where P is the period of pulsations, $\Delta P$ - average absolute deviation of the individual period from the mean value, $\Delta E$ – intervals of the cycles during which the star maintained the constant period. The main conclusion: the value of $\Delta E$ statistically decreases with the increase of the period and the stars of shorter period retain a period of constant for a larger number of cycles.

Continuing the tradition of studies of long-period variable stars in the Astronomical observatory of the Odessa University, we try to conduct a detailed classification type stars Mira-type, semi-regular variable stars and similar, find out the location of symbiotic stars among other long-period variables in their evolution to the planetary nebulae, using compiled material from other authors and their own observations. The modern research of long-period variable stars in the Observatory of the Odessa National University is conducted in several directions.
First of all, there is no special distinction between Mira-type stars and red semi-regular variables. These stars are investigated by the same methods, despite the light curves of semi-regular variables are relatively noisy due to smaller amplitudes of pulsation variability. And the same statistical processing is applied to the long-period components of symbiotic systems, where they dominate in the light curves. Last we believe it is possible, as in several works, such as Whitelock (1987), to note that the symbiotic Mira-type stars are more similar to the Mira than to other symbiotic stars.

For the research, we have used observations from the databases of the French Association of Variable Stars Observers (AFOEV), Variable Stars Observers League of Japan (VSOLJ) and recently the American Association of Variable Stars observers (AAVSO), which allow to study the variability of these stars during the time interval up to 100 years.

A review on the evolutionary properties of long-periodic variables was presented by Kudashkina (2003).

Considering the discontinuity of observations, before the approximation using the trigonometric polynomial, the original data were "cleaned". Points that differ by more than $2^m$ from sure observations, were thrown out. The remaining points were approximated using the "running parabola", then thrown out those points that differ from the approximation by more than $1^m.2$. Details of the method of processing the original data can be found in the paper by Andronov (1987).

Observation interval is typically of about 70 years. Some studies have also used the data from the AAVSO (Mattei, 1979), the processing of which is described in the article by Kudashkina and Andronov (1996). For semi-regular variable stars, the original observations were treated the same way as for the Mira (described above), but light curve was approximated by a running sine (Andronov and Chinarova, 2013). Other methods were described by (Andronov and Chinarova, 2012).

**The first direction.**

The analysis of more than 60 correlations was obtained earlier by Kudashkina and Andronov (1992). Presumably, the stars are divided into basic, intermediate and carbon groups.

We considered some correlations between the photometric parameters of the mean light curves, as such as the slope ascending and descending branches versus period and other.

The parameter $r_1$ is the first harmonic contributed the light curve (the semi-amplitude of the wave with a frequency corresponding to the sinusoidal light curve).

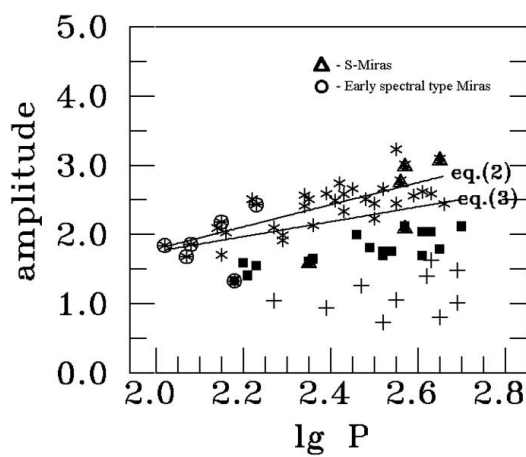

Fig. 1. The relationship "semi-amplitude $r_1$ - period": $r_1 = (1.59\pm0.22)\cdot\lg P - (1.40\pm0.53)$ (2) – only asterix; $r_1 = (1.07\pm0.32)\cdot\lg P - (0.39+0.77)$ (3) – asterix and squares. The crosses mark small-amplitude stars, which were not taken into account.

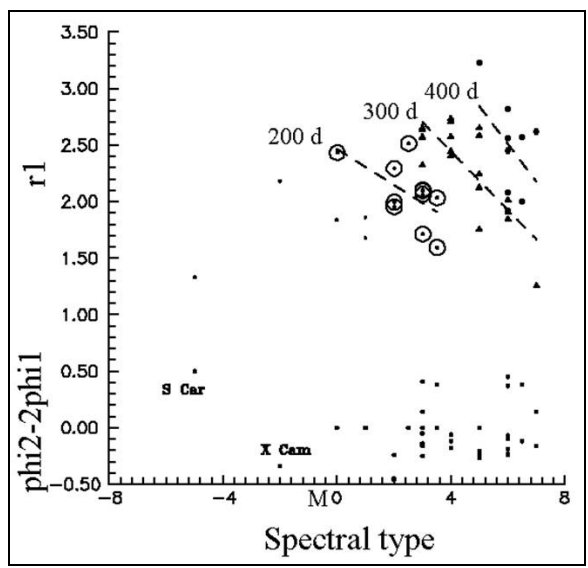

Fig.2. $r_1$ - amplitude of the main wave of the light curve, phi$_2$-2phi$_1$ - difference of the phase of the main and second waves at the light curve. The circles mark the stars with symmetric light curves.

Thus, we examined a possibility to use the photometric parameters of visual light curves of long-periodic stars to derive fundamental characteristics: the stellar population, to which the star belongs, evolutionary stages, pulsating modes etc.

We have concluded: (1) zirconium stars (S-Miras) are the final stage of the Mira-type stars on the AGB. In the Fig. 1, the stars marked with asterixes have initial masses larger then the stars marked with squares. The latter, perhaps, have a metal deficit; (2) the variations of total shape of the light curve versus period caused by only variable of the slope of the ascending branch; (3) for zirconium and oxygen Mira-type stars of the late spectral subclasses, the asymmetry of the light curve is constant and does not depend on the period; (4) for stars with the same periods, the increase of the spectral subclass is accompanied with a decrease of the amplitude of the main wave of the light curve (Fig. 2); (5) the sequence of small-amplitude Mira-type stars exists and includes mainly C-rich stars. Perhaps, exist two sorts (of C- rich stars or of small amplitude Miras): "intrinsic" – the stars, which have passed the stage of the "dredge-up" and "false" – the stars, which belong to the binary systems. The enrichment by C takes place as a result of the mass transfer and the uncovering more deep shells, by analogy with the Barium-stars (Mennessier, 1997).

Most of the stars is ahead of the light curve in the phase of $r_2$ to $r_1$, i.e. the maxima of the sine components at the main and double frequency occur prior to the maximum of the total approximation of the light curve. All stars with delay in the phase of $phi_2$ to $phi_1$, have the humps on the light curves.

Perhaps, for the stars X Aur and R Vir the half period is given in "General Catalogue of Variable Stars" (Samus et al. 2014).

The described features of 8 stars are listed in table 1 (Kudashkina, 1999).

Table 1.

| Star | Features | Spectral type |
|------|----------|---------------|
| W And | S – type, $\Delta m \sim 11^m$ | S6.1e – S9.2e |
| RT Cyg | Supergiant | M2e – M8.8eIb |
| X Aur | Pulsates in 1 overtone? | M3e – M6e |
| T Cam | S – type | S4.7e – S8.5e |
| S UMi | Double maximum? | M6e – M9e |
| T Cas | Double maximum | M6e – M9.0e |
| R Aql | Helium – flash stage | M5e – M9e |
| U CMi | Double maximum | M4e |

**The second direction.**

The changes of individual characteristics of the light curves with time are studied for the long-period variable stars. The characteristics of individual cycles for the 53 stars obtained using methods of "running" and "asymptotic" parabolae are listed by Marsakova and Andronov (2006).

The changes of these characteristics with time are studied. The secular (progressive) period changes for 4 stars, which presumably undergo the helium flash were studied by Marsakova & Andronov (2000a). The characteristics of period changes were investigated. T UMi shows a clear increase in the slope *dm/dt* of the branches of the light curve, R Aql and R Hya show a strong amplitude decrease, mainly due to brightening of minima. If you place these stars in the sequence of changes in the amplitude and slope branches of the light curve, this sequence is in a good agreement with the position of the stars at different stages of the helium flash on existing models (Wood & Zarro, 1981).

For other Mira-type stars also the studies of secular variations of light curve parameters such as period, amplitude, mean brightness have been made. Several types of period variations for Mira-type variables were separated (Marsakova, 2014b).

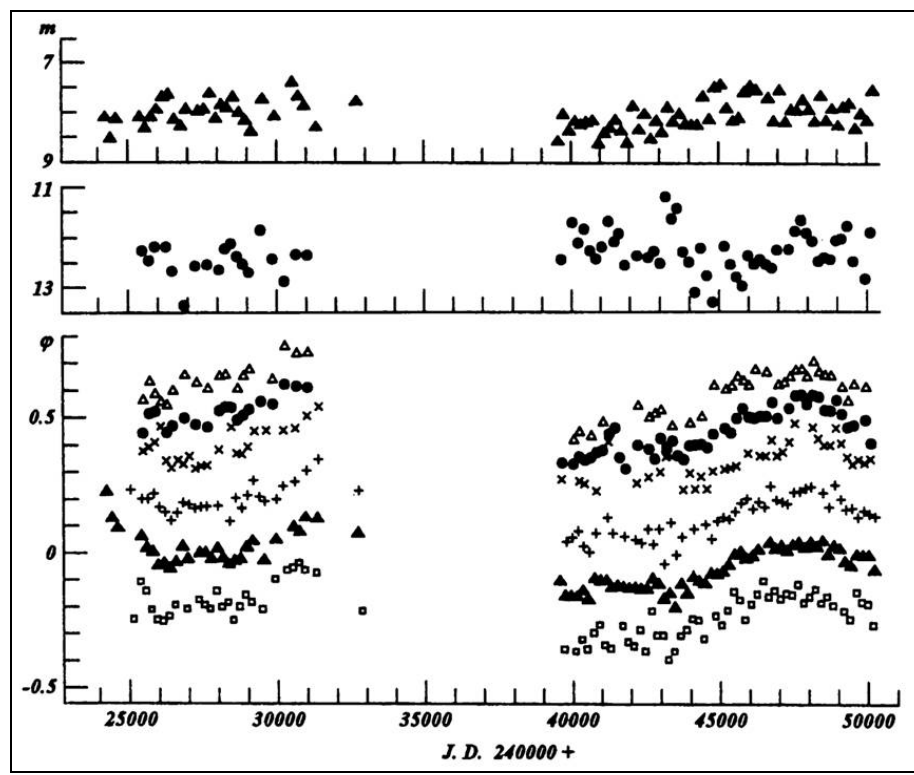

Fig. 3. Variation of magnitudes of maxima and minima and the phases of individual cycle points with time for W Lyr (Marsakova & Andronov, 1997)

In a few stars were discovered cyclic changes of periods with characteristic times of 17000-20000 days (Marsakova V.I. & Andronov I.L., 2013). At T Cep and U UMi, these changes are accompanied by changes in the amplitude. For T Cep (Marsakova V.I. & Andronov I.L., 2000b) its uniform of a number of observations have been able to conduct cross-correlation analysis and to establish that the amplitude minimum, when the hump on the ascending line of the most striking, and between these characteristics there is no time shift. Between periods and amplitudes were obtained shift 3 cycles.

The changes of the characteristics of individual cycles of the stars of spectral classes C and S (carbon and zirconium) are studied (Marsakova & Andronov, 1999). Characteristic features are: high amplitude S – stars and a low amplitude in C-stars (and the average amplitudes of the light curves for 80 years often have values, typical for semi-regular stars), strong variability of light curves, the C-stars, the lack of systematic changes of the period and amplitude. Also, in the C stars are observed significant changes in average luminosity. The assumption was made about the relationship of these changes to dust shells from these stars.

It is shown that classification of long-periodic variables can be based on the character and numeric characteristics of the light curve parameters variations (Marsakova and Andronov, 2006; Marsakova, 2014a)

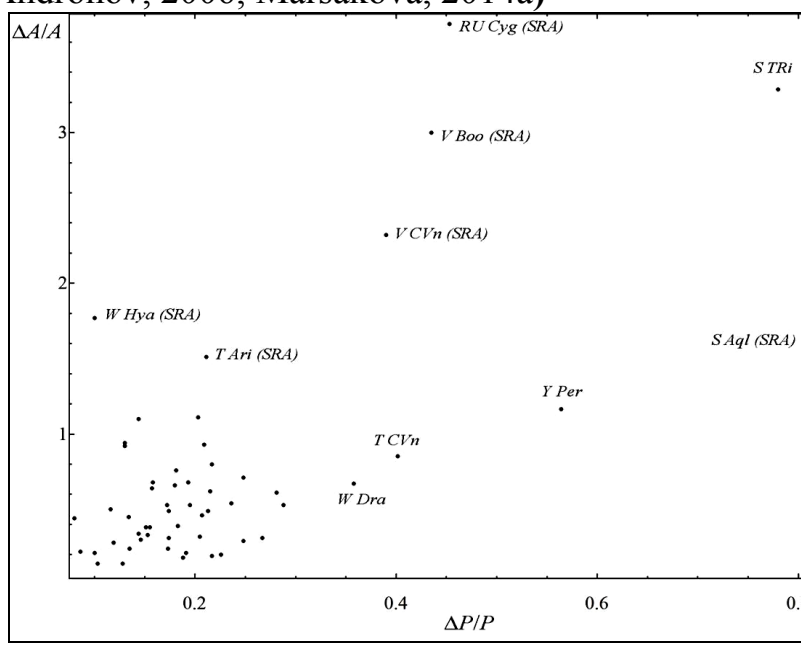

Fig.4. Diagram "Relative scatter of individual cycle amplitudes versus relative scatter of periods" show separation of semiregular and multiperiodic variables from the other Miras (Marsakova, 2014a).

The relationship between the values that characterize changes in the characteristics of individual cycles with other parameters of stars was analyzed. In particular, we studied a number of charts such as: spectral class – relative variation of the amplitude, spectral class – relative scatter of the mean (over the individual cycle) brightness, relative scatter of amplitudes of individual cycles – the average amplitude of the light curve, relative scatter of amplitudes of individual cycle versus relative scatter of periods.

**The third direction**

The variability of 10 objects that are transitional between Mira-type stars and semi-regular variables have been considered.

The stars are divided into two groups: I - those who differ from the Mira – type only by smaller amplitudes (the formal limiting value in the "General Catalogue of Variable Stars" (Samus et al. 2014) is $2^m$), II - have observed transitions to full – blown semi-regular behavior in certain time intervals.

Based on these charts and character of changes of the period, amplitude and medium (over individual period) brightness, one can make the following additions to the classification of these stars. (1) Mira-type stars and semi-regular stars of spectral class C can be assigned to one close group of variable stars. As typically semi–regular, only 4 stars can be classified, two stars were assigned to the intermediate class between the Mira – type and semi-regular variables (Y Per – the Mira – type star and S Aql – SRa). Among a class of titanium stars, were also allocated two stars that were attributed to the so – called "small – amplitude Mira – type stars" that have an amplitude of less than $2^m$, but show features of variability typical for the Mira-type stars. (2) The type SRc is to be divided into at least two subtypes. SRca – supergiants with multi-periodic pulsations and regular light curve of Mira – type, sometimes intervals disturbed by switching modes of pulsations (S Per) and SRcb - supergiants with quasi-periodic light curve and intervals of constancy of brightness (PZ Cas). (3) The group of Mira – type and semiregular variables with similar periodicity (multiperiodicty) is analysed. They have periods of 230–260 days and 140–150 days and show intervals of periodical (Mira-type) variability with relatively high amplitude and "semi-regular" (SR-type) small-amplitude oscillations. Results of periodogram analysis are represented in the paper (Marsakova & Andronov 2013).

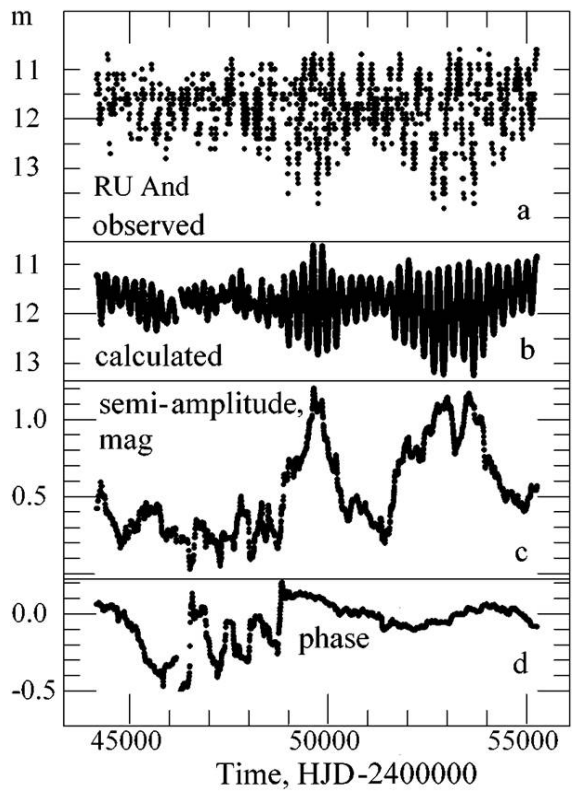

Fig. 5. Light curve of RU And: observations (up), running sine fit, semi-amplitude and phase (bottom) (Chinarova, 2010).

Our analysis in all three studying directions allowed to classify the stars according to the next subtypes:

1) Mira – type stars belonging to the helium – flash stage (TPAGB) – $\langle A_{group}\rangle = 4^m.68$, M-spectral type. The period decreases abruptly and then it increases again, but this process is short time, and the period decreases smoothly. The amplitude changes similarly.

2) The typically Mira-type stars M-spectral type, $\langle A_{group}\rangle = 4^m.36$. The light curves are regular, amplitudes are more $2^m$. Perhaps, the amplitude increases for several stars.

3) The small-amplitude Mira-type stars - $\langle A_{group}\rangle = 1^m.70$, M-spectral type. The amplitudes are less than $2^m$. All parameters are typical for Miras. As it was mentioned above some of them may be companions of binary systems.

4) Mira-type stars of the spectral type S - $\langle A_{group}\rangle = 5^m.44$ They belong to the final stage of the Mira-type stars on the AGB.

5) Regular long-period variables C-spectral type. The amplitudes are small, light curves are more noisy, instable, chaotic, have humps or double maxima. For classification of C-stars as Mira of SR the same criteria (as for M-stars) can not be applied (amplitude, regularity).

6) Transient type stars (between Miras and SR) - $\langle A_{group}\rangle = 1^m.34$. The switches between regular and semi-regular periodicities have been observed. These stars drop out at our diagrams.

7) Semi-regular variables M-spectral type - $\langle A_{group}\rangle = 0^m.84$. The amplitude often decrease during the time; the mean brightness changes during the time.

Thus, in traditional classification of long-period variables, the basic criteria are mean amplitude and the regularity of light curves. We propose the parameters and diagrams, which can be used for more detailed classification. These parameters are based on the character and range of the variations characteristics of individual and mean light curves.

*Acknowledgements.* The authors are thankful to Dr. Bogdan Wszołek and AJD for hospitality. For this work, we have used the international databases of the AFOEV, VSOLJ, AAVSO. The work is included in the "Inter-Longitude Astronomy" (Andronov et al., 2010) and "Ukrainian Virtual Observatory" (Vavilova et al., 2012) projects.